\documentclass[3p]{elsarticle} 

\usepackage{url,hyperref,microtype,subcaption}
\usepackage{acro}
\usepackage{siunitx}
\usepackage{multirow}
\usepackage{amssymb}
\usepackage{amsmath}
\usepackage{latexsym}
\usepackage{graphicx}
\usepackage[onehalfspacing]{setspace}

\def\firstAuthorLast{Turan {et~al.}} 

\def\Address{$^{1}$ Faculty of Aerospace Engineering, Department of Space Engineering, Delft University of Technology, Delft, 2629 HS, The Netherlands }
\def\corrAuthor{Corresponding Author}

\def\corrEmail{e.turan@tudelft.nl}

\DeclareAcronym{LiAISON}{short = LiAISON, long = Linked Autonomous Interplanetary Satellite Orbit Navigation}
\DeclareAcronym{BER}{short = BER, long = Bit Error Rate}
\DeclareAcronym{CRTBP}{short = CRTBP, long = Circular Restricted Three-body Problem}
\DeclareAcronym{LU}{short = LU, long = Length Unit}
\DeclareAcronym{TU}{short = TU, long = Time Unit}
\DeclareAcronym{EKF}{short = EKF, long =  Extended Kalman Filter}
\DeclareAcronym{STM}{short = STM, long = State Transition Matrix}
\DeclareAcronym{RMS}{short = RMS, long = Root Mean Square}
\DeclareAcronym{LOS}{short = LOS, long = Line-of-Sight}
\DeclareAcronym{OM}{short = OM, long = Observability Matrix}
\DeclareAcronym{SVD}{short = SVD, long = Singular Value Decomposition}
\DeclareAcronym{SNR}{short = SNR, long = Signal-to-Noise}
\DeclareAcronym{FIM}{short = FIM, long = Fisher Information Matrix}
\DeclareAcronym{CRLB}{short = CRLB, long = Cramér–Rao Lower Bound}
\DeclareAcronym{SKF}{short = SKF, long = Schmidt-Kalman filter}
\DeclareAcronym{LRO}{short = LRO, long = Lunar Reconnaissance Orbiter}
\DeclareAcronym{ETP}{short = ETP, long = Europa Tomography Probe}
\DeclareAcronym{ISL}{short = ISL, long = Inter-Satellite Link}
\DeclareAcronym{DTE}{short = DTE, long = Direct-to-Earth Link}
\DeclareAcronym{LUMIO}{short = LUMIO, long = Lunar Meteoroid Impact Observer}
\DeclareAcronym{EML2}{short = EML$_2$, long = Earth-Moon L$_2$}
\DeclareAcronym{EML1}{short = EML$_1$, long = Earth-Moon L$_1$}
\DeclareAcronym{OD}{short = OD, long = orbit determination}
\DeclareAcronym{EIRP}{short = EIRP, long = Equivalent Isotropic Radiated Power}
\DeclareAcronym{SRP}{short = SRP, long = Solar Radiation Pressure}
\DeclareAcronym{ECI}{short = ECI, long = Earth-Centered Inertial}
\DeclareAcronym{MCI}{short = MCI, long = Moon-Centered Inertial}
\DeclareAcronym{CKF}{short = CKF, long = Consider-Kalman Filter}
\DeclareAcronym{LPF}{short = LPF, long = Lunar Pathfinder }
\DeclareAcronym{PN}{short = PN, long = Pseudo-Noise }
\DeclareAcronym{S/C}{short = S/C, long = Spacecraft }
\makeatletter
\def\ps@pprintTitle{%
 \let\@oddhead\@empty
 \let\@evenhead\@empty
 \def\@oddfoot{}%
 \let\@evenfoot\@oddfoot}
\makeatother
\begin{document}

\begin{frontmatter}

\title{Autonomous Crosslink Radionavigation for a Lunar CubeSat Mission}%

\author{Erdem Turan\corref{cor1}}
\cortext[cor1]{Corresponding author}
  \ead{e.turan@tudelft.nl}
\author{Stefano Speretta}
\author{Eberhard Gill}

\address{Delft University of Technology, Faculty of Aerospace Engineering, Kluyverweg 1, 2629 HS, Delft, the Netherlands}

\begin{abstract}
This study presents an autonomous orbit determination system based on crosslink radiometric measurements applied to a future lunar CubeSat mission to clearly highlight its advantages with respect to existing ground-based navigation strategies. This work is based on the \ac{LiAISON} method which provides an autonomous navigation solution solely using satellite-to-satellite measurements, such as range and/or range-rate, to estimate absolute spacecraft states when at least one of the involved spacecraft has an orbit with a unique size, shape, and orientation. The lunar vicinity is a perfect candidate for this type of application due to the asymmetrical gravity field: the selected lunar mission, an \ac{EML2} Halo orbiter, has an inter-satellite link between a lunar elliptical frozen orbiter. Simulation results show that, even in case of high-measurement errors (in the order of \SI{100}{m}, $1\sigma$), the navigation filter estimates the true states of spacecraft at \ac{EML2} with an error in the order of \SI{500}{m} for position, and \SI{2}{mm/s} for velocity, respectively and the elliptical lunar frozen orbiter states can be estimated in the order of \SI{100}{m} for position and \SI{1}{cm/s} for velocity, respectively. This study shows that range-only measurements provide better state estimation than range-rate-only measurements for this specific situation. Different bias handling strategies are also investigated. It has been found that even a less accurate ranging method, such as data-aided ranging, provides a sufficient orbit determination solution. This would simplify the communication system design for the selected CubeSat mission. The most observable states are found to be position states of the lunar orbiter via the observability analysis. In addition, the best tracking windows are also investigated for the selected mission scenario. 
\end{abstract}

\end{frontmatter}

\section{Introduction}

There has been an increasing interest in small satellites for lunar missions lately, forming almost 40\% of all planned deep space small satellite missions \citep{Turan2022}. In these missions, the baseline option for orbit determination is, in general, ground-based radiometric navigation. However, ground-based tracking could be expensive and limited considering crowded ground networks. In addition, small satellite missions are expected to be low-cost and there are also limitations from small satellite themselves such as on-board power for communication. Autonomous navigation, on the other hand, could be a possible approach for these lunar missions. Until now, various autonomous navigation methods have been proposed and implemented. One of them, called \ac{LiAISON}, uses solely satellite-to-satellite observations, such as range or range-rate, to estimate the absolute states of the involved satellites when at least one of them has an orbit with unique size, shape, and orientation \citep{hillthesis, Hill2007,Hill2008}. The characteristics of the acceleration function determine whether inter-satellite range or range-rate measurements can be used alone to estimate the absolute and relative \ac{S/C} states. Considering the asymmetric gravity field in the cislunar vicinity, it is possible to build such an autonomous navigation system. Up to now, various studies have presented the capabilities of \ac{LiAISON} over the past decade in lunar and deep space mission studies \citep{Leonard2012,Wang2019,leonard2015thesis,Fujimoto2016, turanaero2022}. 

This navigation method will be tested via the link between \ac{LRO} and the CAPSTONE CubeSat soon \cite{Cheetham2020}. However, many more missions targeted to the \ac{EML2} could benefit from such technique. One of these, \ac{LUMIO} is a CubeSat at \ac{EML2} designed to observe, quantify, and characterize the meteoroid impacts by detecting their flashes on the Lunar farside, to provide global information on the Lunar Meteoroid Environment \citep{Cervone2022,topputo2021current,Speretta2018,speretta2022lumio}. The baseline navigation strategy, as for almost all small satellites targeting this orbit, is ground-based radiometric. Beside this, \ac{LUMIO} also includes an inter-satellite link to a larger Lunar orbiter for telecommand purposes but not for navigation. Having an inter-satellite link provides an opportunity to investigate the performances of autonomous navigation via crosslink radiometric measurements, potentially extending the mission possibilities with this new technique.

This study presents the autonomous navigation performances of the selected mission scenario: \ac{LUMIO}, via the link between \ac{LPF} and the \ac{LUMIO} CubeSat. A simulation-based analysis will determine the achievable orbit determination accuracy considering realistic radio frequency measurement errors derived from the Phase-A study inter-satellite link design, which was not designed to perform navigation. In the following sections, at first, the selected mission scenario is presented and then, dynamical models are provided. Orbit determination models are introduced including the observability analysis. Thereafter, the navigation simulation setup and results are presented. Finally, conclusions are drawn. 

\section{A Lunar CubeSat scenario}

For this study, the \ac{LUMIO} mission has been selected. It  features a 12U CubeSat in a halo orbit at the \ac{EML2} to observe, quantify, and characterize meteoroid impacts on the Lunar farside by detecting their flashes \citep{topputo2021current, Cervone2022}. The mission aims at determining the spatial and temporal characteristics of meteoroids impacting the Lunar surface to characterize their flux. The operative orbit for \ac{LUMIO} has been selected as a quasi-periodic halo orbit with a Jacobi constant $C_j=3.09$ \citep{Cervone2022}. The \ac{LUMIO} Phase-A study has been completed in March 2021 while the mission’s Phase B is expected to start soon \citep{speretta2022lumio}. 

The radiometric navigation system uses existing hardware of the communication system featuring a combination of \ac{ISL} and \ac{DTE} links. The latter has been designed to provide payload data downlink, ranging and tracking in nominal conditions. Based on the \ac{OD} analysis given in \citep{speretta2021}, ground-based radiometric navigation via Cebreros, ESTRACK or the Sardinia Deep Space Antenna for 3 hours per track following a 7 + 7 + 14 days scheme meets the \ac{OD} requirements of \SI{1}{km} and \SI{1}{cm/s} position and velocity accuracy, respectively. In this study the same \ac{OD} requirements have been considered for an autonomous navigation scenario. The \ac{ISL}, on the other hand, has been designed to provide a redundant commanding link without involving a dedicated deep-space class ground station but reusing commercial resources. Such link provides optimal performances in terms of visibility, despite the fact that data rates are quite limited. The SSTL \ac{LPF} spacecraft has been considered as relay satellite. Depending on the relative distance, between a minimum of \SI{31000}{km} and a maximum of \SI{89000}{km}, data rates are expected in the order of \SI{0.5}{kbps}-\SI{4}{kbps} based on S-band, \SI{9}{dBW} \ac{EIRP} link (also including a \SI{3}{dB} safety margin \cite{speretta2022lumio}). 

\section{Autonomous Radiometric Navigation}

This study investigates the \ac{LiAISON} orbit determination performances for the \ac{LUMIO} mission: autonomous orbit determination requires absolute position and velocity estimates without any ground-based observation. This requires that the \ac{S/C} states, obtained from inter-satellite observations, must be observable. In the two body problem, relative range or range-rate measurements do not provide full state estimation due to the rank deficiency: absolute orientations of the orbital planes are not observable, which is related to the symmetrical gravity field. However, in an asymmetrical gravity field, spacecraft-to-spacecraft tracking provides absolute state estimation. The \ac{LiAISON} orbit determination method uses solely inter-satellite measurements to estimate absolute \ac{S/C} states of the involved \ac{S/C} when at least one of them has an orbit with a unique size, shape and orientation \citep{hillthesis, Hill2007, Hill2008}. Basically, if one of the \ac{S/C} has an unique orbit, it is possible to estimate absolute states of all involved spacecrafts. This study derives inter-satellite observations, namely range and range-rate, from radio frequency measurements. 

Inter-satellite radiometric measurements can be collected with various methods. Range observations, for example, can be collected via phase or time-based measurements: the phase shift on a ranging signal at the receiver with respect to a transmitter provides an accurate ranging solution (in the order of \SI{1}{m}). This still requires the phase ambiguity to be solved in case the ranging signal period exceeds the round-trip light time. There are lots of common standards used for this purpose such as pseudo noise/sequential tone-based ranging \citep{Turan2022}. However, small satellites, in general, have limited on-board power available for data transfer. If a small satellite requires ranging for its navigation, such signal reduces the power available for telemetry in case both signals are modulated at the same time window. On the other hand, timing exchange between satellites provides also a ranging solution. Basically, in order to deal with on-board power limitations, time-derived ranging methods are a possible options for small satellites. These methods are not quite accurate (in the order of \SI{150}{m} at \SI{10}{kbps}) but they can provide sufficient ranging solutions to meet the navigation requirements. Because time-derived ranging methods are data-rate dependent, high-data transfer between satellites is beneficial. Another measurement type is range-rate and this can be derived from the Doppler shift between satellites but, again, this measurement type suffers from on-board limitations. In this study, high and low measurement ranging errors derived from the conventional (radiometric) and time-derived ranging methods are considered for the selected mission scenario.

\section{Orbit Determination Models}

This section presents the orbit determination models used in this study. Dynamical, measurement and estimation models are given in the following subsections.

\subsection{Dynamical Model}

The dynamical model used in this study is formulated as \ac{CRTBP}. This is simple but accurate enough for this type of application, considering the mission required position accuracy. In \cite{hillthesis}, the \ac{LiAISON} \ac{OD} performances remain in the same order of magnitude for various force models. In addition to \ac{CRTBP}, in the end, high-fidelity dynamic simulation models results will also be given for the selected measurement configuration to have a more realistic analysis and to compare results obtained from \ac{CRTBP} dynamics.

\begin{figure}[h!]
\begin{center}
\includegraphics[width=10cm]{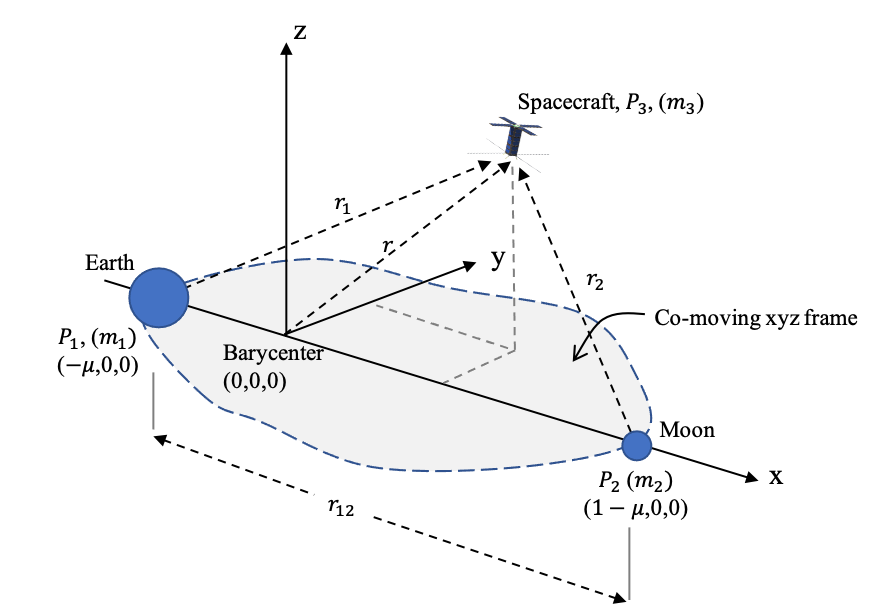}
\end{center}
\caption{ Circular Restricted Three-body Problem: Earth-Moon co-moving reference frame}\label{fig:CRTBP}
\end{figure}

The \ac{CRTBP} assumes there are two massive bodies, Earth ($P_1$) with mass $m_1$ and Moon ($P_2$) with mass $m_2$ in this case, moving under their mutual gravitation in a circular orbit around each other with a radius $r_{12}$, \citep{Curtis2020}. Considering a non-inertial, co-moving reference frame (see Figure~\ref{fig:CRTBP}) with its origin at the barycenter of the two bodies, the positive $x-$axis points from the barycenter to $P_2$. The positive $y-$axis is parallel to the velocity vector of $P_2$ and the $z-$axis is perpendicular to the orbital plane. 

Considering a third body of mass $m_3$ with $m_3 \ll m_1$ and $m_3 \ll m_2$, it cannot impact the motion of primary bodies, $P_1$ and $P_2$. The equations of motion for the \ac{CRTBP} are \citep{hillthesis}:

\begin{equation}
\ddot{x}-2\dot{y}=x-(1-\mu)\frac{x+\mu}{r_1^3}-\mu\frac{x+\mu-1}{r_2^3}
\end{equation}
\begin{equation}
    \ddot{y}+2\dot{x}=(1-\frac{1-\mu}{r_1^3}-\frac{\mu}{r_2^3})y 
\end{equation}
\begin{equation}
\ddot{z}=(\frac{\mu-1}{r_1^3}-\frac{\mu}{r_2^3})z
\end{equation}

where $r_1=\sqrt{(x+\mu)^2 + y^2 + z^2}$ and $r_2=\sqrt{(x+\mu-1)^2 + y^2 + z^2}$. For the Earth-Moon system, the gravitational parameter $\mu$ is  0.01215, the normalized time $t^*$ \SI{4.343}{days}, and the normalized length $l^*$ \SI{384747.96}{km}, respectively. The transformation from the rotating frame to the inertial frame can be done as follows \cite{amanda2010}:
\begin{equation}
    X_{inertial}=T X_{rot}
\end{equation}
where
\begin{equation}
T=\begin{bmatrix}
L & 0_{3\times 3} \\ 
\dot{L} & L
\end{bmatrix}
\end{equation}
and
\begin{equation}
L=\begin{bmatrix}
\cos(t) & -\sin(t) & 0\\ 
\sin(t) & \cos(t) & 0 \\ 
0 & 0 & 1
\end{bmatrix}, \hspace{1cm} \dot{L}=\begin{bmatrix}
-\sin(t) & -\cos(t) & 0\\ 
\cos(t) & -\sin(t) & 0 \\ 
0 & 0 & 1
\end{bmatrix}
\end{equation}
where $X_{rot}$ represents the non-dimensional Earth-Moon barycentric coordinates, $X_{inert}$ represents the inertial state centered on the barycenter. It is assumed that two frames coincide at the initial time $t_0=0$.  A rotating position vector of the \ac{S/C} in the Moon-centered frame is as follows:
\begin{equation}
    X_{rot,Moon}=X_{rot}-\begin{bmatrix}
1-\mu\\ 
0\\ 
0
\end{bmatrix}
\end{equation}
and in the Earth-centered frame:
\begin{equation}
    X_{rot,Earth}=X_{rot}-\begin{bmatrix}
\mu\\ 
0\\ 
0
\end{bmatrix}
\end{equation}
Based on the equations given above, states can be transformed from the Earth-Moon barycenteric rotating frame to the primary-centered, either Earth or Moon, inertial frame. In the study, the Lunar orbiter's initial states are expressed in the \ac{MCI} frame, a coordinate transformation is needed to convert to the non-dimensional Earth-Moon barycentric frame. Basically, these steps are followed in reverse order. 

In high-fidelity dynamical simulations, in addition to gravitational acceleration due to Earth and Moon (treated as point masses in the \ac{ECI} J2000 frame), the gravitational acceleration due to Sun treated as a point mass and acceleration due to \ac{SRP} with a spherical model have been used. 

\subsection{Measurement Model}

In this study, the autonomous navigation method uses solely inter-satellite measurements and the observables are collected via radiometric measurements. The first radiometric data type used is range, which can be derived from either conventional ranging methods or data-aided ranging. A conventional ranging method using is pseudo-random noise has been selected and, considering a non-coherent transponder with a pseudo-noise square wave shaped ranging signal, and a chip tracking loop, the following one-way ranging error (1$\sigma$) would be expected \cite{book2014pseudo}:
\begin{equation}
    \label{rsigma}
    \sigma _{\rho_{PN}} =\frac{c}{8 f_{rc}}\sqrt{\frac{B_{L}}{(P_{RC}/N_{0})}}
\end{equation}

and the range bias due to a chip rate mismatch:
\begin{equation}
    \label{rbias}
    \rho _{\text{bias}}=\frac{c\Delta f_{chip}T}{4f_{chip}}
\end{equation}

with $c$ the speed of light, $f_{rc}$ the frequency of the ranging clock component, $B_{L}$ one-sided loop noise bandwidth, $P_{RC}$ power of the ranging clock component, $T$ integration time, $N_{0}$ one-side noise power spectral density, $\Delta f_{chip}$ the difference in frequency between the received chip rate and the local chip rate. 

\begin{figure}[h!]
\begin{center}
\includegraphics[width=10cm]{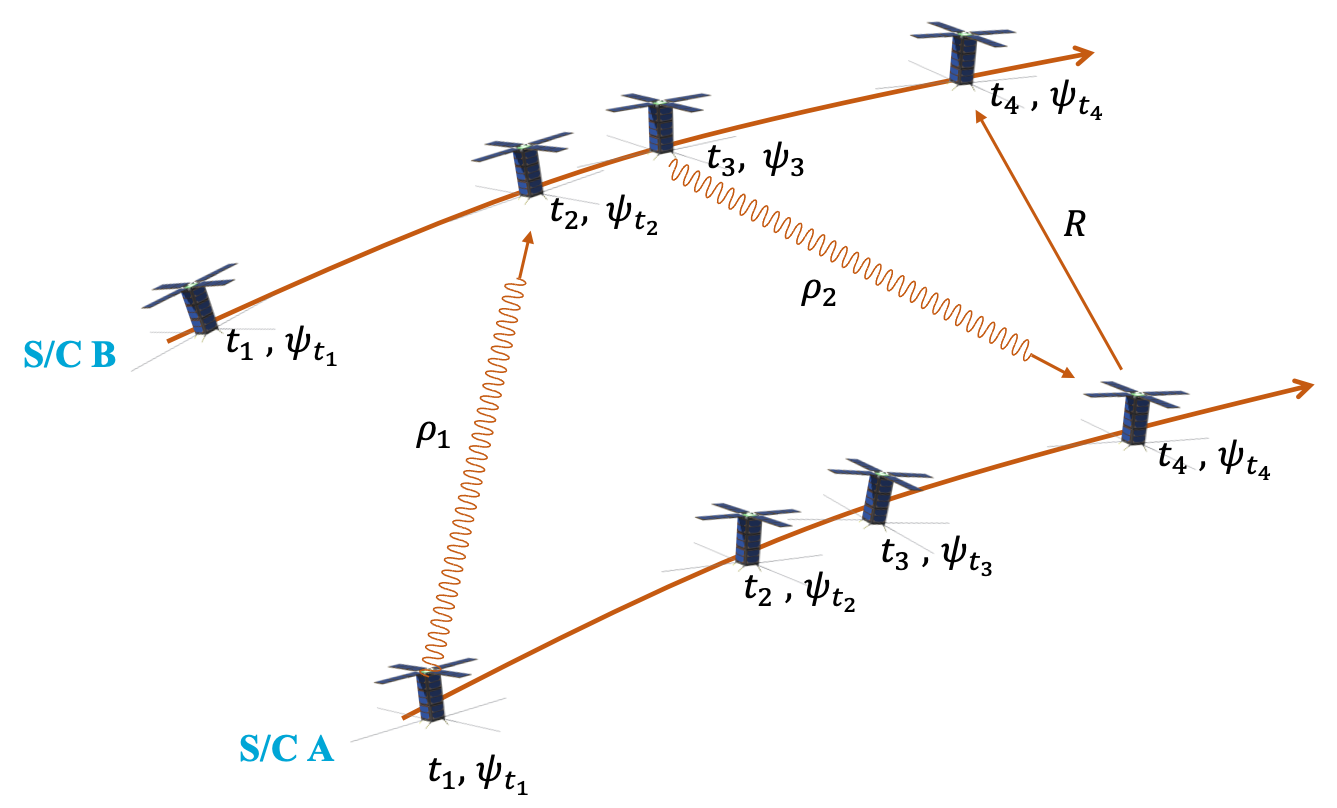}
\end{center}
\caption{ Time exchange between satellites for the purpose of time-derived ranging. $t_i$ represents time, $\psi_i$ represents onboard clock states at $t_i$} \label{fig:ISL}
\end{figure}

The round-trip light time can also be measured via time transfer between satellites. This process requires four successive time stamps to be obtained which represent the time of transmission and reception of both \ac{S/C} (see Figure~\ref{fig:ISL}). If the timing is measured in units of telemetry/telecommand symbols, instead of directly in seconds, the performance of time-derived ranging (one-way) would be:
\begin{equation}
    \label{tdrange}
    \sigma _{\rho_{TD}} = \frac{4 \, c \, T_{sd}^{2}}{\pi \, T_{l} \, E_{S}/N_{0}}   
\end{equation}

where $T_{sd}$ is the symbol duration, $T_{l}$ the correlator integration time and $E_{S}/N_{0}$ the symbol-to-noise ratio. Note that both equations \ref{rsigma} and \ref{tdrange} represent one-way ranging performance and this requires calculations for both uplink and downlink. 

The measurement error for two-way Doppler due to thermal noise, influencing range-rate observations, can be given as follows \cite{dsnranging}:
\begin{equation}
    \label{rvsigma}
    \sigma _{V} = \frac{c}{2 \sqrt{2} \pi f_{c} T}\sqrt{\frac{1}{\rho_{L}}+\frac{G^{2} B_{L}}{(P_{C}/N_{0})}}
\end{equation}
where $f_{c}$ the downlink carrier frequency, $P_{C}/N_{0}$ uplink carrier power to noise spectral density ratio, $\rho_{L}$ the downlink carrier loop signal-to-noise ratio, $G$ the turn-around ratio. Doppler data noise can be expressed by the phase noise, $\sigma_\varphi$, in radians and converted to range-rate noise as follows \citep{Montenbruck2000}:

\begin{equation}
\label{eqnsigmarangerate}
    \sigma_{\dot{\rho}}=\frac{\sqrt2c}{2\ G\ f_t\ t_c}\frac{\sigma_\varphi}{2\pi}
\end{equation}

In this study, the formation includes two \ac{S/C}, a lunar orbiter, \ac{LPF}, and \ac{EML2} Halo orbiter, \ac{LUMIO}. The state vector being estimated consists of the position and velocity components of both \ac{S/C} is as follows:
\begin{equation}
\mathbf{X}=\begin{bmatrix}
x_1 \hspace{2mm} y_1 \hspace{2mm} z_1 \hspace{2mm} {\dot{x}}_1 \hspace{2mm} {\dot{y}}_1 \hspace{2mm} {\dot{z}}_1 \hspace{2mm} x_2 \hspace{2mm} y_2 \hspace{2mm} z_2 \hspace{2mm} {\dot{x}}_2 \hspace{2mm} {\dot{y}}_2 \hspace{2mm} {\dot{z}}_2
\end{bmatrix}^T
\end{equation}
The measurement model in this paper, referred as the pseudo-range, involves the geometric range, the overall clock bias, and other error sources. The two-way ranging measurement concept can be seen in Figure~\ref{fig:ISL}. The geometric range is given as follows:
\begin{equation}
    R=\frac{1}{2}\,c\,(t_4 - t_1 ) + \Delta \rho
\end{equation}
By ignoring the light-time correction, and by assuming the speed of light is greater than the \ac{S/C} relative velocity, $c\gg v$, the geometric range can be modeles as
\begin{equation}
    R= \sqrt{(x_1 -x_2)^{2}+(y_1 -y_2)^{2}+(z_1 -z_2)^{2}}
\end{equation}
where $x_i, y_i$ and $z_i$ represents the position components of \ac{S/C}, $i=1,2$ states and the pseudo range observations can be modeled as
\begin{equation}
    \rho = R + c(\psi _{t_4}-\psi _{t_1})+c\,(\Delta_{tx}+\Delta_{rx})+c\,\Delta_{trx}+\rho_{\text{noise}}
\end{equation}
\begin{equation}
    \rho = \sqrt{(\mathbf{r}_1-\mathbf{r}_2)\cdot (\mathbf{r}_1-\mathbf{r}_2)}+\rho_{\text{bias}} + \rho_{\text{noise}}
\end{equation}
where $\psi _{t_4}$ and $\psi _{t_1}$ are the clock states at $t_4$ and $t_1$ respectively. The transponder transmit and receive line delays are $\Delta_{tx}$ and $\Delta_{rx}$, respectively and $\Delta_{trx}$ is the line delay on the \ac{S/C} transponding the ranging signal. All these terms are combined as $\rho_{\text{bias}}$ and $\rho_{\text{noise}}$ representing the un-modelled statistical error sources.

The range rate measurements, $\dot{\rho}$, can be modelled as:
\begin{equation}
    \dot{\rho}=\frac{\boldsymbol{\rho}\cdot{\dot{\boldsymbol{\rho}}}}{\rho}
\end{equation}
\begin{equation}
    \dot{\rho}=\frac{\left(x_1-x_2\right)\left({\dot{x}}_1-{\dot{x}}_2\right)+\ \left(y_1-y_2\right)\left({\dot{y}}_1-{\dot{y}}_2\right)+\left(z_1-z_2\right)\left({\dot{z}}_1-{\dot{z}}_2\right)}{\sqrt{\left(x_1-x_2\right)^2+\left(y_1-y_2\right)^2+\left(z_1-z_2\right)^2}} + \dot{\rho}_{\text{bias}} + \dot{\rho}_{\text{noise}}
\end{equation}

$\rho_{\text{noise}}$ and $\dot{\rho}_{\text{noise}}$ are calculated based on equations given in \ref{rsigma}, \ref{tdrange}, and \ref{rvsigma}.

\subsection{Estimation Model}
In this study, an \ac{EKF} is adopted as a common method used in real-time navigation. The \ac{EKF} consists of a prediction and a correction step: in the former, predicted state and error covariance $\bar{P}$ are \citep{schutz2004statistical}:
\begin{equation}
    \dot{\mathbf{X}} = \boldsymbol{F}({\mathbf{X}},t), \hspace{5mm} {\mathbf{X}}(t_{k-1})=\hat{\mathbf{X}}_{k-1}
\end{equation}
\begin{equation}
\label{eqnPmeas}
\boldsymbol{\bar{P}}_k = \boldsymbol{\Phi}(t_k,t_{k-1})\boldsymbol{P}_{k-1}\boldsymbol{\Phi}^T(t_k,t_{k-1}) + \boldsymbol{Q}
\end{equation}

where $\boldsymbol{\Phi}(t_k,t_{k-1})$ is the state transition matrix from $t_{k-1}$ to $t_{k}$ and $\boldsymbol{Q}$ is the process noise matrix. The correction step is:

\begin{equation}
    \boldsymbol{K}_k=\boldsymbol{\bar{P}}_k \boldsymbol{\tilde H}_k^T [\boldsymbol{\tilde H}_k \boldsymbol{\bar{P}}_k \boldsymbol{\tilde H}_k^T +\boldsymbol{W}_k]^{-1}
\end{equation}
\begin{equation}
    \hat{\mathbf{X}}_k = {\mathbf{X}}_k + \boldsymbol{K}_k [\mathbf{y}_k - \boldsymbol{\tilde H}_k {\mathbf{X}}_k ]
\end{equation}
\begin{equation}
\label{eqnPtime}
    \boldsymbol{P}_k =[\boldsymbol{I}-\boldsymbol{K}_k \boldsymbol{\tilde H}_k] \boldsymbol{\bar{P}}_k
\end{equation}

where $\bold{\hat{X}}$ is the state estimate, $\boldsymbol{K}$ is the Kalman gain, $\boldsymbol{\tilde H}$ is the measurement sensitivity, $\boldsymbol{P}$ is the error covariance estimate, and $\boldsymbol{W}$ is the state noise compensation matrix. 

In this study, the state noise compensation is introduced by $Q$ which can be constructed for each \ac{S/C} as follows \citep{Hill2008}:
\begin{equation}
    Q_m=\begin{bmatrix}
\frac{\Delta t^4 \sigma^2_i}{3} & 0 & 0 & \frac{\Delta t^3 \sigma^2_i}{2} & 0 & 0 \\ 
 0 & \frac{\Delta t^4 \sigma^2_i}{3} & 0 & 0 & \frac{\Delta t^3 \sigma^2_i}{2} & 0 \\ 
 0 & 0 & \frac{\Delta t^4 \sigma^2_i}{3} & 0 & 0 & \frac{\Delta t^3 \sigma^2_i}{2} \\ 
\frac{\Delta t^3 \sigma^2_i}{2} & 0 & 0 & \Delta t^2 \sigma^2_i & 0 & 0 \\ 
0 & \frac{\Delta t^3 \sigma^2_i}{2} & 0 & 0 & \Delta t^2 \sigma^2_i & 0 \\ 
0 & 0 & \frac{\Delta t^3 \sigma^2_i}{2} & 0 & 0 & \Delta t^2 \sigma^2_i
\end{bmatrix}, \hspace{1cm} m=1,2
\end{equation}
The measurement bias is expected to affect the the navigation system performances: in general, this can either be neglected or estimated by including dynamic or measurement model parameters into the state vector. Another approach would be to assume its \textit{a priori} estimate and associated covariance matrix are known. In this study, all three cases (neglected bias, estimated bias and considered bias) were investigated. In case of bias estimation, the estimated state vector must be expanded with a bias component, $\rho_{\text{bias}}$. This requires $\boldsymbol{\tilde H}$ and $\boldsymbol{\Phi}$  to expand as follows:
\begin{equation}
    \boldsymbol{\tilde H}=\begin{bmatrix}
\frac{\partial \mathbf{\rho}}{\partial \mathbf{r_1}} & \frac{\partial \mathbf{\rho}}{\partial \mathbf{v_1}} & \frac{\partial \mathbf{\rho}}{\partial \mathbf{r_2}} & \frac{\partial \mathbf{\rho}}{\partial \mathbf{v_2}} & 1
\end{bmatrix} \hspace{1cm} \boldsymbol{\Phi}(t_k,t_{k-1})=\begin{bmatrix}
\boldsymbol{\Phi}_\mathbf{Y} & 0\\ 
0 & 1
\end{bmatrix}
\end{equation}
In a similar way, the clock drift can be estimated by expanding $\boldsymbol{\tilde H}$ and $\boldsymbol{\Phi}$ with $(t-t_0)$ representing the step size and $1$ representing unchanged clock parameter during the time update process, respectively. However, expanding them into the state vector may affect the performances by reducing observability of the navigation system. That's why this study also investigated the considered bias case by implementing a sequential consider filter (known as \ac{CKF}, \ac{SKF}). Here, the specific time invariant measurement bias case is studied assuming that consider parameter, $b_k$, is constant for all $k$ and the bias covariance matrix, $B_k$, is time invariant. \ac{CKF} is slightly different than \ac{EKF} and requires to implement the following equation into the time update:
\begin{equation}
\boldsymbol{\bar{C}}_k = \boldsymbol{\Phi}(t_k,t_{k-1})\boldsymbol{C}_{k-1}
\end{equation}
and changes in measurement update: 
\begin{equation}
        \boldsymbol{K}_k=(\boldsymbol{\bar{P}}_k \boldsymbol{\tilde H}_k^T + \boldsymbol{\bar{C}}_k \boldsymbol{N}_k^T) \boldsymbol{\Omega}_k^{-1}
\end{equation}
\begin{equation}
    \hat{\mathbf{X}}_k = {\mathbf{X}}_k + \boldsymbol{K}_k [\mathbf{y}_k - \boldsymbol{\tilde H}_k {\mathbf{X}}_k -{\mathbf{N}}_k {\mathbf{b}}_0]
\end{equation}
\begin{equation}
        \boldsymbol{P}_k =[\boldsymbol{I}-\boldsymbol{K}_k \boldsymbol{\tilde H}_k] \boldsymbol{\bar{P}}_k - \boldsymbol{K}_k \boldsymbol{\Omega}_k \boldsymbol{\bar{C}}_k^T
\end{equation}
\begin{equation}
    \boldsymbol{C}_k = \boldsymbol{\bar{C}}_k - \boldsymbol{K}_k (\boldsymbol{H}_k \boldsymbol{\bar{C}}_k + \boldsymbol{N}_k \mathbf{B}_0)
\end{equation}
where
\begin{equation}
    \boldsymbol{\Omega}_k = \boldsymbol{\tilde H}_k \boldsymbol{\bar{P}}_k \boldsymbol{\tilde H}_k^T + \boldsymbol{N}_k \boldsymbol{\bar{C}}_k^T \boldsymbol{\tilde H}_k^T + \boldsymbol{\tilde H}_k \boldsymbol{\bar{C}}_k \boldsymbol{N}_k^T + \boldsymbol{N}_k \mathbf{B}_0 \boldsymbol{N}_k^T + \mathbf{W}_k
\end{equation}

where $\boldsymbol{C}_k $ is the cross-covariance matrix, $\boldsymbol{B}_0$ is the bias covariance, $\boldsymbol{b}_0$ \textit{a priori} the measurement biases estimate, and $\boldsymbol{N}_k$ the  measurement bias vector sensitivity matrix. Note that the \ac{CKF} turns into \ac{EKF} in case of zero bias.

\subsection{Observability}

The observability analysis is used to relate \ac{OD} performance and observation data: in this study, the degree of the system observability is used to evaluate the estimation performances. For this purpose, the observability Gramian is used as follows:

\begin{equation}
\label{Neqn}
    \boldsymbol{N}=\sum_{k=1}^{l}\boldsymbol{\Phi}(t_k,t_0)^T \boldsymbol{\tilde{H}}^T_k\boldsymbol{\tilde{H}}_k \boldsymbol{\Phi}(t_k,t_0) 
\end{equation}

The observability can be assessed via \ac{SVD} of the observability Gramian. In general, two metrics are used for this purpose: the condition number, which is the ratio of the largest singular value to the smallest one, and unobservability index, which is the reciprocal of the smallest local singular value. Most and least observable states can also be derived from unitary matrices via \ac{SVD}.

\section{Navigation Simulations}

This section presents the orbit determination results for the selected mission scenario. The simulation setup is given first and results are given thereafter in the corresponding subsections.

\subsection{Simulation setup}

\begin{figure}[h!]
\begin{center}
\includegraphics[width=15cm]{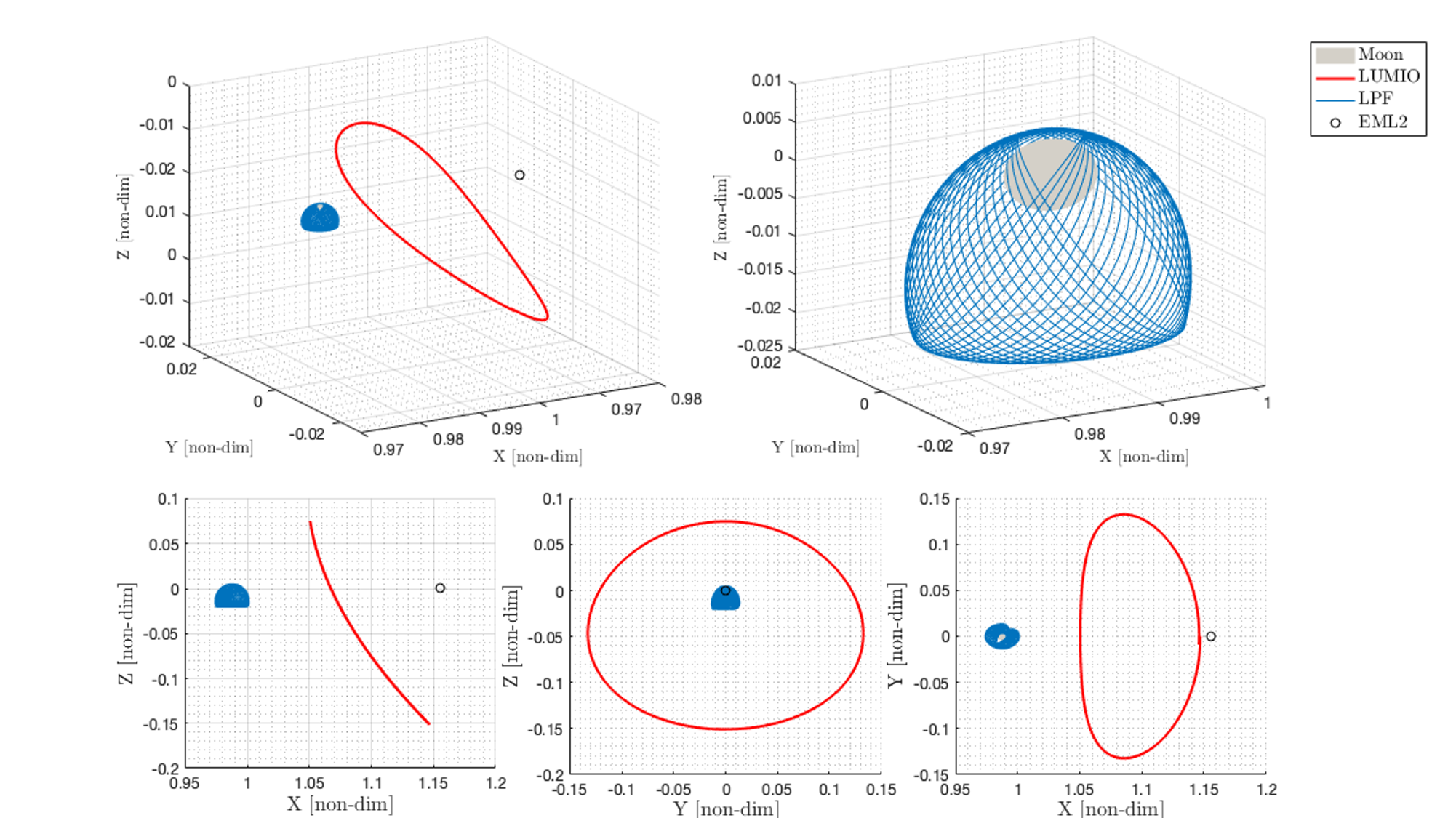}
\end{center}
\caption{Trajectories of both S/C for 14 days in the non-dimensional Earth-Moon barycentric frame}\label{fig:Traj1}
\end{figure}

The selected mission scenario consists of two \ac{S/C}, the \ac{LPF} and \ac{LUMIO}, at the Elliptical Lunar Frozen orbit and \ac{EML2} Halo orbit \citep{scotti2022, Cervone2022}. The simulation duration is set to be 14 days. For the the N-body orbital dynamics based analysis, the simulation start time and duration are set to be 18 April 2024, 21:00:00 UTC. The initial states of \ac{LPF} expressed in \ac{MCI} are listed in Table~\ref{LPF}. \ac{LUMIO} has a quasi-halo orbit at \ac{EML2} with a Jacobi energy of $C_j=3.09$. For this study, due to the duration of the simulation, a similar southern Halo orbit (with the same Jacobi energy) has been used in \ac{CRTBP}. The initial conditions can be found in Table~\ref{ICTable} including the \ac{LPF} converted states from \ac{MCI} to the non-dimensional Earth-Moon barycentric frame. Corresponding trajectories of both \ac{S/C} for 14~days can be seen in Figure~\ref{fig:Traj1}.

\begin{table}[h!]
\centering
\caption{\ac{LPF} initial states expressed in \ac{MCI} \citep{scotti2022}}
\label{LPF}
\begin{tabular}{ll} 
\hline\hline
\textbf{Parameter} & \textbf{Value} \\ 
\hline
Semi-major axis & \SI{5737.4}{km} \\
Eccentricity & 0.61 \\
Inclination & \SI{57.83}{deg} \\
RAAN & \SI{61.55}{deg} \\
Argument of periselene & \SI{90}{deg} \\
True anomaly at Epoch & \SI{0}{deg} \\
\hline\hline
\end{tabular}
\end{table}

\begin{table}[h!]
\centering
\caption{Initial states used in the simulations (expressed in non-dimensional Earth-Moon barycentric frame)}
\label{ICTable}
\begin{tabular}{lcc} 
\hline\hline
\multicolumn{1}{c}{S/C} & Position & Velocity \\ 
\hline
\ac{LPF} & \begin{tabular}[c]{@{}c@{}}0.98512134\\0.00147649\\0.00492546\end{tabular} & \begin{tabular}[c]{@{}c@{}}-0.87329730\\-1.61190048\\0\end{tabular} \\
LUMIO & \begin{tabular}[c]{@{}c@{}}1.1473302\\0\\-0.15142308\end{tabular} & \begin{tabular}[c]{@{}c@{}}0\\-0.21994554\\0\end{tabular} \\
\hline\hline
\end{tabular}
\end{table}

In the mission scenario, two-different measurement error cases have been simulated which were representing the high-accuracy conventional pseudo-noise ranging method and the time-derived ranging method. Inter-satellite ranging is relaying on  the communication system and the link budget for \ac{LUMIO} is presented in Table~\ref{radioparam}. Inter-satellite range measurement parameters for both conventional and time-derived methods are given in Table~\ref{radioparam}. In this table, different from the existing link budget, the uplink data-rate and thus symbol duration is increased in consideration of a higher gain antenna configuration or higher transmit power. This is reasonable because the existing \ac{S/C} configuration did not take into account the inter-satellite link based autonomous navigation (the existing configuration would give \SI{2700}{m} $1\sigma$ ranging accuracy which is not sufficient for the mission). Based on these assumptions, range and range-rate measurement performances can be seen in Table~\ref{radioparam}.

\begin{table}
\centering
\caption{Assumptions for \ac{LUMIO} (the link budget is given based on the maximum inter-satellite distance \citep{speretta2022lumio}. Range measurement parameters are for conventional and time-derived methods and measurement errors are $1\sigma$, two-way)}
\label{radioparam}
\begin{tabular}{llcc} 
\hline\hline
 & \multicolumn{1}{c}{\textbf{Parameter}} & \multicolumn{2}{c}{\textbf{Value }} \\ 
\hline\hline
 &  & Downlink & Uplink \\
\multirow{8}{*}{\begin{tabular}[c]{@{}l@{}}Inter-Satellite\\Link Budget\end{tabular}} & Frequency, $f$ & \SI{2200}{MHz} & \SI{2100}{MHz} \\
 & TX power, $P_t$ & \SI{3}{dBW} & \SI{3}{dBW} \\
 & TX path losses, $L_t$ & \SI{1}{dB} & \SI{1}{dB} \\
 & TX antenna gain, $G_t$ & \SI{6.5}{dBi} & \SI{23.6}{dBi} \\
 & Polarisation loss, $L_p$ & \SI{0.5}{dB} & \SI{0.5}{dB} \\
 & Data rate & \SI{4000}{bps} & \SI{500}{bps} \\
 & Required $E_b/N_0$ & \SI{2.5}{dB} & \SI{2.5}{dB} \\
 & Link Margin & \SI{3}{dB} & \SI{3}{dB} \\ 
\hline
\multirow{10}{*}{\begin{tabular}[c]{@{}l@{}}Radiometric\\Measurement\\Parameters\end{tabular}} & Symbol rate, $1/T_{sd}$ & \SI{4000}{sps} & \SI{2700}{sps} \\
 & Correletor integration time, $T_l$ & \multicolumn{2}{c}{\SI{0.5}{s}} \\
 & Symbol-to-noise ratio, $E_s / N_0$ & \multicolumn{2}{c}{\SI{-1}{dB}} \\
 & Modulation & \multicolumn{2}{c}{BPSK} \\
 & Transponding ratio, $G$ & \multicolumn{2}{c}{$1$} \\
 & Range clock frequency, $f_{rc}$ & \multicolumn{2}{c}{\SI{1}{MHz} } \\
 & Ranging code & \multicolumn{2}{c}{T2B} \\
 & Ranging clock power over & \multicolumn{2}{c}{\multirow{2}{*}{\SI{25}{dBHz}}} \\
 & noise spectral density, $P_{rc}/N_0$ & \multicolumn{2}{c}{} \\
 & Loop Bandwidth, $B_L$ & \multicolumn{2}{c}{\SI{1}{Hz}} \\
 & Chip rate difference, $\Delta f_{chip}$ & \multicolumn{2}{c}{\SI{100}{Hz}} \\ 
\hline
\multirow{3}{*}{\begin{tabular}[c]{@{}l@{}}Measurement\\Errors\end{tabular}} & Conventional PN ranging error & \multicolumn{2}{c}{\SI{2.98}{m}} \\
 & Time-derived ranging error & \multicolumn{2}{c}{\SI{102.44}{m}} \\
 & Range-rate error & \multicolumn{2}{c}{\SI{0.97}{mm/s}} \\
\hline\hline
\end{tabular}
\end{table}

During the simulations, in each $k$th time step, the \ac{RMS} error for the $N$th case of the Monte Carlo simulation has been calculated by using following
\begin{equation}
\label{eqnmontecarlo}
    {RMSE}_{k}=\sqrt{\frac{1}{N}\sum_{i=1}^{N}(x_{i,k}-\hat{x}_{i,k})^{2}}
\end{equation}
where $x_{i,k}$ and $\hat{x}_{i,k}$ are $i$th component of state vector and its estimate respectively. Parameters used in simulations can be found in Table~\ref{tablefilterparam}. Regarding filter uncertainty and initial errors, a more detailed analysis has been done during the \ac{LUMIO} Phase-A design study, previously. A ground-based tracking for 7 hours between 18 April 2021 14:00:00 and 21:00:00 UTC from the Sardinia Deep Space Antenna based on range and range-rate measurements (measurement errors of \SI{1}{m} range and \SI{0.33}{mm/s} range-rate with a measurement bias of \SI{2.5}{m}) would give \SI{0.11}{km} and \SI{0.95}{cm/s} position and velocity errors with $1\sigma$ uncertainty of \SI{1.65}{km} and \SI{4.7}{cm/s}, respectively (Estimation time Epoch: 18 April 2021 21:00:00 UTC with Batch-Least squares). Basically, this single tracking session estimation can be used to initialize the autonomous navigation system on-board. A ground-based state estimation has not been done for \ac{LPF}. However, it is assumed that ground-based state estimation results for \ac{LPF} is the same order of magnitude. In brief, initial parameters given in Table~\ref{tablefilterparam} are sensible. 
\begin{table}[h!]
\centering
\caption{Parameters used in simulations}
\label{tablefilterparam}
\begin{tabular}{lc} 
\hline\hline
\multicolumn{1}{c}{\textbf{Parameter}} & \textbf{Value} \\ 
\hline
Position uncertainty, ($x_1,y_1,z_1,x_2,y_2,z_2$), $1\sigma$~ & 1km \\
Velocity uncertainty, ($\dot{x}_1,\dot{y}_1,\dot{z}_1,\dot{x}_2,\dot{y}_2,\dot{z}_2$), $1\sigma$~ & 1cm/s \\
Initial position error, ($x_1,y_1,z_1,x_2,y_2,z_2$) & 500 m \\
Initial velocity error, ($\dot{x}_1,\dot{y}_1,\dot{z}_1,\dot{x}_2,\dot{y}_2,\dot{z}_2$) & 1mm/s \\
Measurement error & See Table~\ref{radioparam} \\
Systematic bias & 10 m \\
\hline\hline
\end{tabular}
\end{table}

As already mentioned, \ac{OD} requirement of  the \ac{EML2} orbiter is \SI{1}{km} for position and \SI{1}{cm/s} for velocity, respectively. This has been taken as baseline goal for this study. The trajectory used in the study is also considered as a true reference, so there is no error in the dynamics. In addition to \ac{CRTBP} dynamical models, the N-body orbital dynamics simulations are also performed. In such model (based on the JPL DE405 ephemeris model), Earth, Moon, and Sun are treated as point masses. Regarding the \ac{SRP} model, \ac{SRP} areas are set to \SI{3}{m^2} and \SI{0.41}{m^2}, and reflectivities are set to 1.8 and 1.08, for \ac{LPF} and \ac{LUMIO}, respectively \citep{scotti2022, sirani2021}. The other settings for the high-fidelity analysis are the same.

\subsection{Results}

This section presents the performance of radiometric autonomous navigation for the selected mission scenario. The effects of measurement accuracy, precision, data type and navigation filter on the \ac{OD} performance have been investigated.

\begin{figure}[h!]
\begin{center}
\includegraphics[width=15cm]{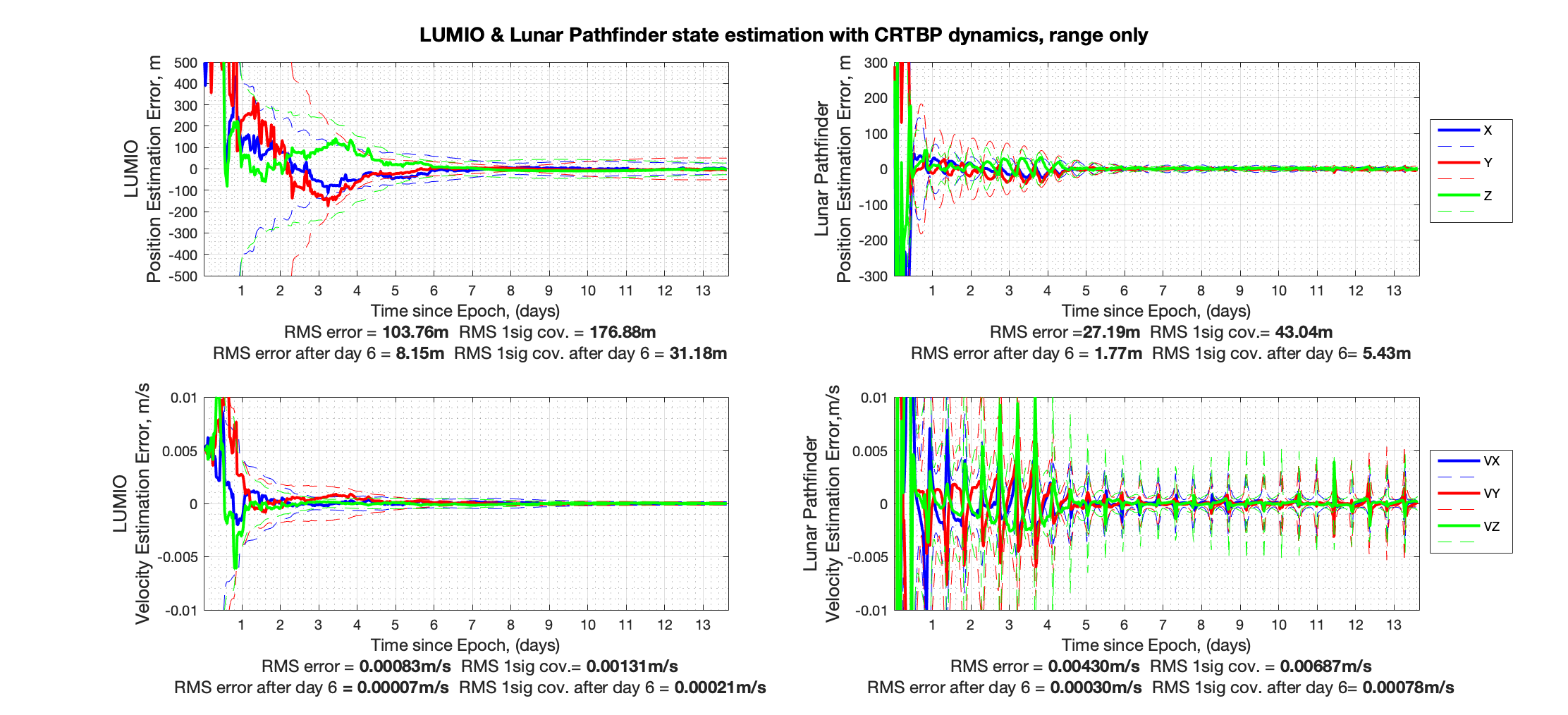}
\end{center}
\caption{State estimation results based on range-only measurements}\label{fig:3m1run}
\end{figure}

\begin{figure}[h!]
\begin{center}
\includegraphics[width=15cm]{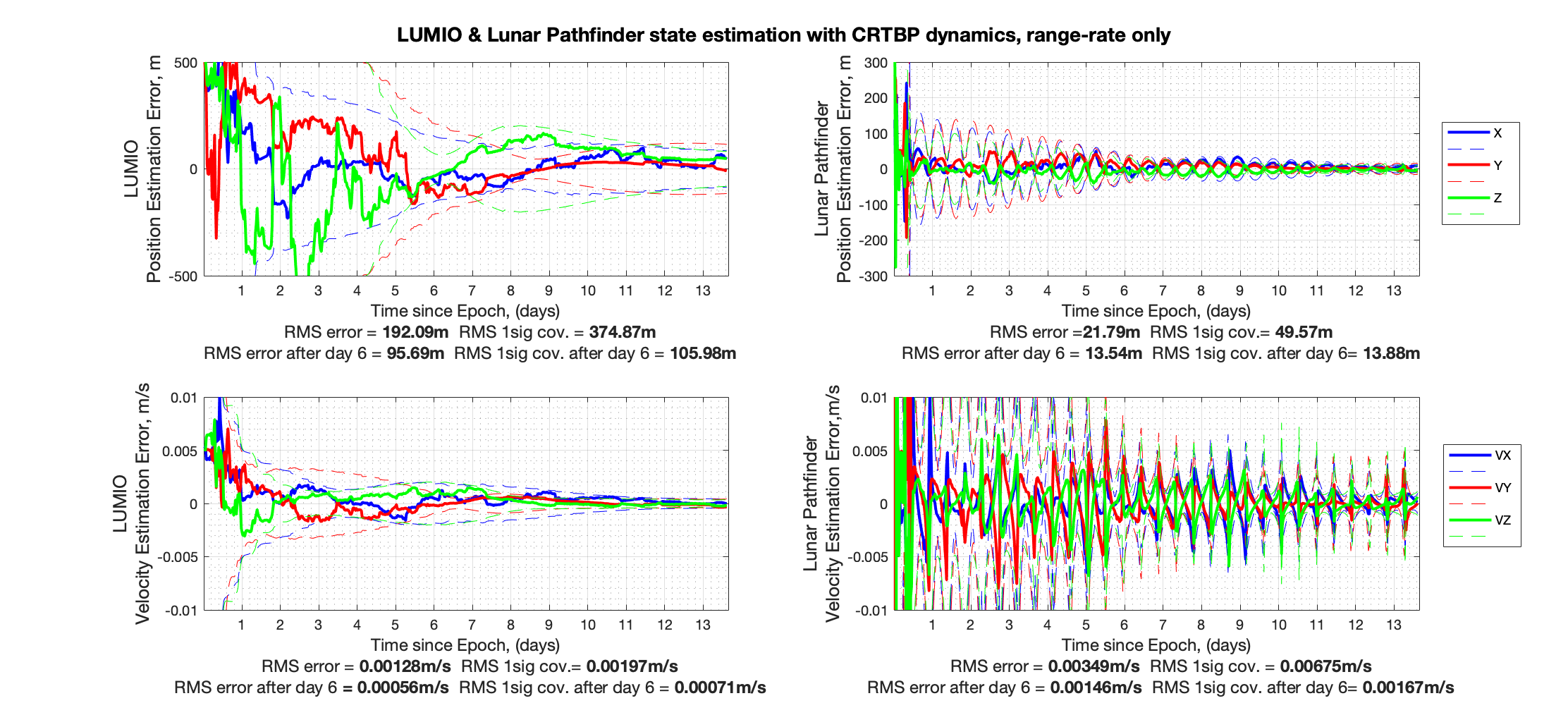}
\end{center}
\caption{State estimation results based on range-rate-only measurements}\label{fig:0_9mms1run}
\end{figure}

At first, the baseline case has been presented to show the autonomous navigation method works in the lunar vicinity. This case is based on the inter-satellite ranging derived from the conventional \ac{PN} method. Based on the settings given in the previous section, the navigation filter estimates the true states of \ac{LUMIO} in the order of \SI{100}{m} for position and \SI{1}{mm/s} for velocity, respectively. \ac{LPF} states are estimated within the order of \SI{10}{m} position and \SI{1}{cm/s} velocity, respectively. Estimation results can be seen in Figure~\ref{fig:3m1run} including \ac{RMS} error and covariance values. As it can be seen, the filter converged after day 6. This is due to the fact that halo orbit has a period of 14 days and the half-orbit is sufficient to fit for the \ac{EML2} orbiter. Fluctuations in the \ac{LPF} state estimation are related to the relative geometry between \ac{S/C}. The position estimation converges when \ac{LPF} approaches the periselene, and diverges when \ac{LPF} approaches the aposelene. It is beneficial for \ac{LPF} position estimation to be performed when the \ac{LPF} is in the high velocity region. In case range-rate measurements are used, instead of range, the filter estimates are not improved (see Figure~\ref{fig:0_9mms1run}). Monte-Carlo simulation results can be seen in Table~\ref{MCtable}.
    
\begin{table}[h!]
\centering
\caption{100-execution Monte-Carlo simulation results (values in parenthesis represents the results after day 6, and values are averaged over two S/C)}
\label{MCtable}
\scalebox{0.90}{\begin{tabular}{lcccc} 
\hline\hline
\multirow{2}{*}{} & \multicolumn{2}{c}{\textbf{Position }} & \multicolumn{2}{c}{\textbf{Velocity }} \\
 & \ac{RMS} Error & \ac{RMS} $1\sigma$ uncertainty & \ac{RMS} Error & \ac{RMS} $1\sigma$ uncertainty \\ 
\hline
Range only & 75.25m (17.07m) & 147.09m (48.03m) & 2.65mm/s (0.51mm/s) & 4.39mm/s (0.93mm/s) \\
Range-rate only & 143.03m (49.44m) & 226.31m (63.72m) & 2.82mm/s (1.01mm/s) & 4.24mm/s (1.20mm/s) \\
\hline\hline
\end{tabular}}
\end{table}

As part of the  analysis, it was determined that the most observable states are in order: $z_2$, $x_1,$ $y_2$, $x_2, \dot{y}_1,$ $z_1, y_1,$ $\dot{x}_1, \dot{z}_1,  \dot{y}_2,  \dot{x}_2,  \dot{z}_2. $ (subindex 1 for \ac{LUMIO} and 2 for \ac{LPF}). Basically, the most and least observable states are the position and velocity components of \ac{LPF}, respectively. The condition numbers are $\SI{2.521e12}{}$ and $\SI{4.324e12}{}$ for range-only and range-rate only case, respectively. This shows that s range-only system has a higher observability than a range-rate only system. However, the range-rate only case provides higher information to the filter on the least observable state: $\dot{z}$ via the lower unobservability index ($0.5829$ for range-rate only, and $\SI{4.595e03}{}$ for range-only). The range-only system also converges faster than the range-rate only system. Overall, range measurements provide better state estimation due to relative geometry and lower measurement error. For this reason, the coming part of the paper continues with range-only measurements.

\begin{figure}[h!]
\begin{center}
\includegraphics[width=9cm]{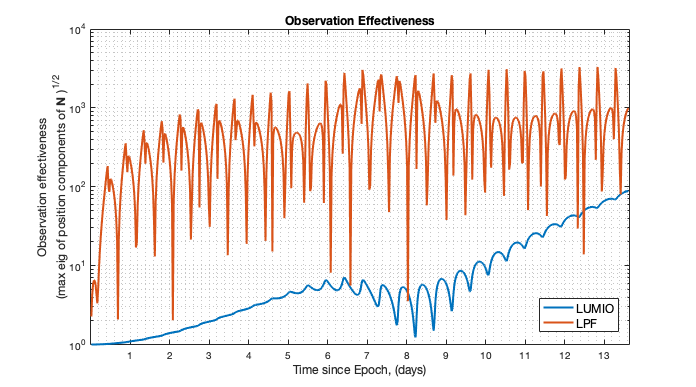}
\end{center}
\caption{Observation effectiveness for both S/C positional states.}\label{fig:obseff}
\end{figure}

The observation effectiveness for the mission scenario has also been investigated. This introduces how much information each measurement provides to the filter. In this case, observation effectiveness on the positional components are given in Figure~\ref{fig:obseff}. As it can be seen, effectiveness is increasing as measurements provide valuable information to the filter, e.g., for almost early 6 days to the position estimation of LUMIO states. However, fluctuations can be seen for the  \ac{LPF} plot which is related to the relative geometry between S/C and after a certain time the effectiveness doesn't increase anymore. Basically, optimal tracking windows can be planned based on these peak and dip periods. 

\begin{figure}[h!]
\begin{center}
\includegraphics[width=9cm]{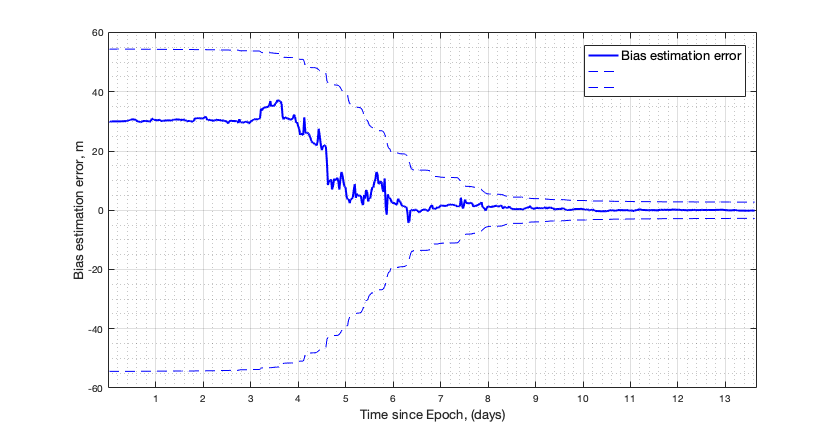}
\end{center}
\caption{Bias estimation error}\label{fig:biasest}
\end{figure}

\begin{figure}[h!]
\begin{center}
\includegraphics[width=16cm]{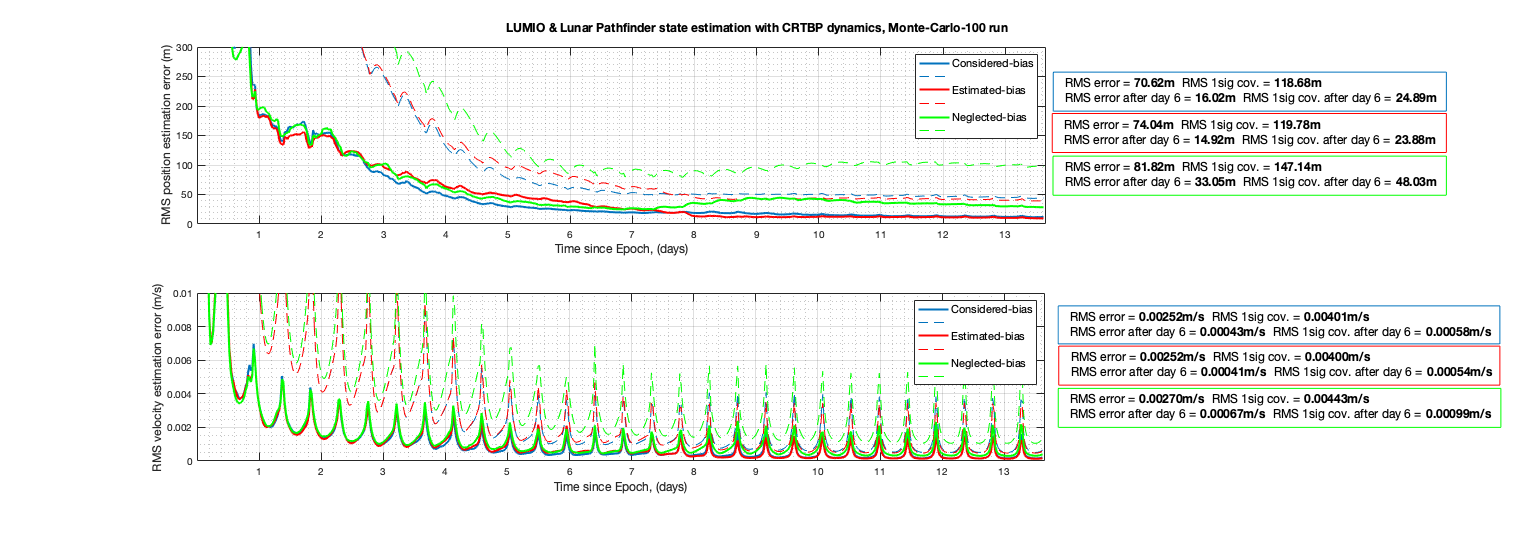}
\end{center}
\caption{Comparison of three different approaches for bias (considered-bias, estimated-bias, and neglected-bias).}\label{fig:biascases}
\end{figure}

In the previous case, the measurement bias has not been included into the simulation to show directly the relation between the data types. However, it is well known, measurement bias affects the navigation system performances and it is important to know how they degrade with a systematic bias. For this purpose, a measurement bias of \SI{10}{m} has been implemented and the same scenario has been re-run with three different cases: considered-bias, estimated-bias, and neglected-bias. In the estimated bias case, estimated state-vector has been expanded with a bias term. As it can be seen from Figure~\ref{fig:biasest}, measurement bias can be estimated along with dynamical states. For all three cases, Monte-Carlo results are visible in Figure~\ref{fig:biascases}. Basically, considered-bias and estimated-bias provides similar performances. However, neglecting the bias increases the state estimation errors.

\begin{figure}[h!]
\begin{center}
\includegraphics[width=16cm]{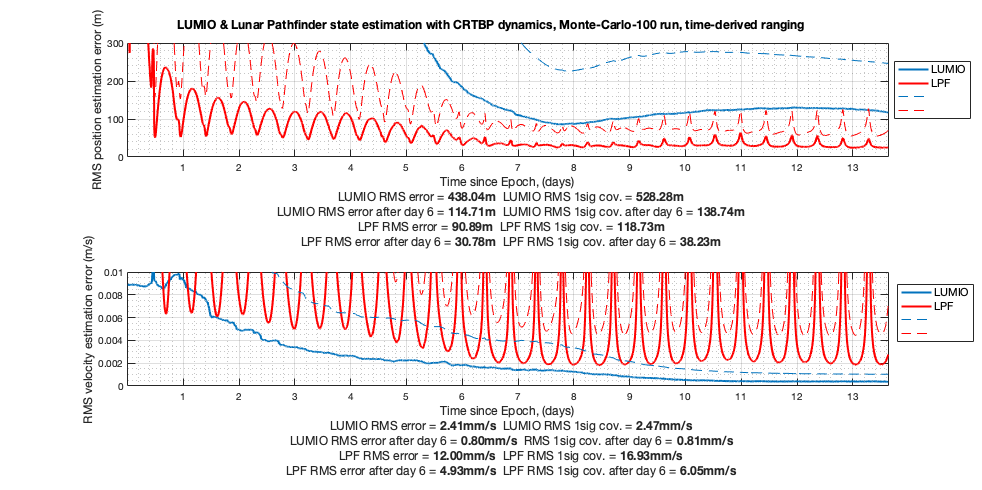}
\end{center}
\caption{State estimation results based on the time-derived ranging method (Monte-Carlo simulation 100-execution)}\label{fig:100mMC}
\end{figure}

In addition to the conventional \ac{PN} ranging, time-derived ranging based simulations have been performed. As already mentioned, modulating the ranging signal reduces the power available for telemetry and thus supported data rates. Considering the link budget, it is not quite easy to perform ranging sessions and telemetry sessions during the same time window. However, time-derived ranging uses telemetry/telecommand signals to estimate the distance between \ac{S/C} without using any additional hardware. This ranging method requires only insertion of the \ac{S/C} clock states into the telemetry and telecommand data frames. Basically, four successive timestamps (reception and transmission time for both \ac{S/C}) are sufficient to calculate the signal round trip light time and time offset. In addition, this method provides a ranging solution along with data-transfer between \ac{S/C} which means there is no need to plan additional ranging sessions and can be used anytime via any type of data transfer between \ac{S/C}. However, this method is not as accurate and this would affect the autonomous navigation performances. In brief, this method would simplify the communication system design and reduce the on-board power required, making this method a perfect candidate for this mission. The time-derived ranging method simulation results are given in Figure~\ref{fig:100mMC}. As it can be noticed, the filter can estimates the true states of \ac{LUMIO} in the order of \SI{140}{m} for position and \SI{1}{mm/s} velocity, respectively. Errors are almost 4 times higher than the conventional ranging method in simulation. However, this still meets the mission  navigation requirements (\SI{1}{km}). This shows, from the navigation performance perspective, that collecting measurements at different time intervals (and thus orbit geometry) has more importance than the absolute measurement errors. As a side note, even in the high measurement error case (time-derived ranging), the system is observable enough to estimate the systematic bias. This only requires more time than in the \ac{PN} case. 

\begin{figure}[h!]
\begin{center}
\includegraphics[width=15cm]{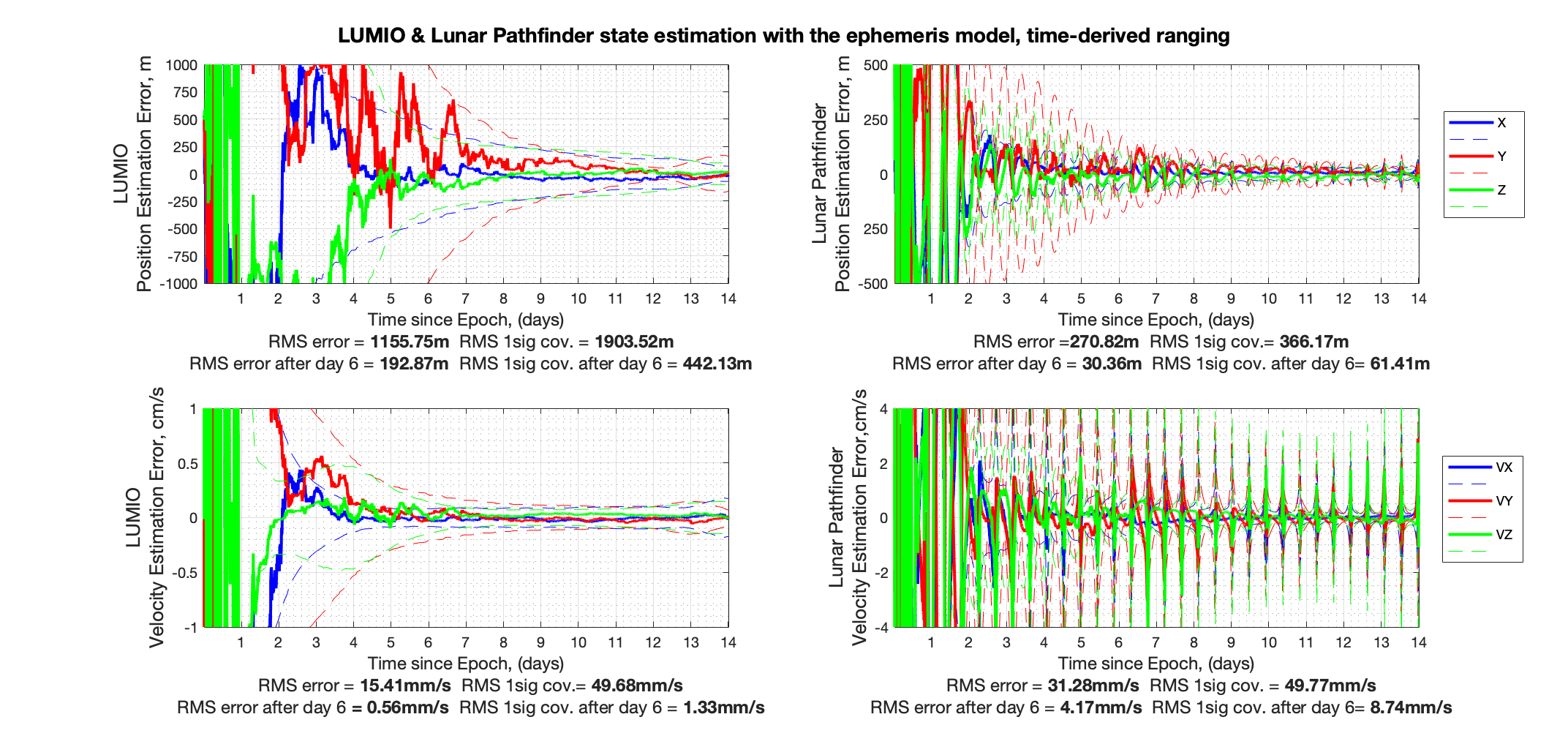}
\end{center}
\caption{State estimation results based on the ephemeris model with time-derived ranging method)}\label{fig:HFMres}
\end{figure}

In the last simulation of the paper, the ephemeris model based simulation results are given. This shows the effects of additional perturbations in the system on the navigation performances. Based on the ephemeris model (including \ac{SRP} and the gravitational acceleration due to Sun), the estimation results are shown in Figure~\ref{fig:HFMres}: position and velocity estimation uncertainty increased up to three times in this case with respect to the \ac{CRTBP} case. However, the \ac{LUMIO} estimation errors are lower than  \SI{500}{m} $1\sigma$ for position and \SI{2}{mm/s} for velocity, respectively, and for \ac{LPF}, $1\sigma$ uncertainties are less than \SI{100}{m} for position and \SI{1}{cm/s} velocity. This would meet the \ac{OD} requirements of \SI{1}{km} position and \SI{1}{cm/s} velocity, respectively. 

\section{Conclusion}

This study showed that the \ac{LiAISON} orbit determination technique could be a possible navigation approach for a proposed mission, \ac{LUMIO}, based on the existing inter-satellite link between \ac{LPF} and the \ac{LUMIO} CubeSat without using any ground based measurements. Simulation results show that the navigation filter estimates the true states of \ac{LUMIO} in the order of \SI{500}{m} for position, and \SI{2}{mm/s} for velocity, respectively. \ac{LPF} states can be estimated in the order of \SI{100}{m} for position and \SI{1}{cm/s} velocity, respectively. Considering the range-only case with $1 \sigma$ error of \SI{2.98}{m} provides better states estimation than the range-rate only case with $1\sigma$ measurement error of \SI{0.97}{mm/s}. The observability analysis showed that the system is observable and found that \ac{LPF}'s position and velocity components are the most and the least observable states, respectively. The best tracking windows are also presented by means of  the observation effectiveness analysis. It has been showed that bias would affect the performances, and it can be estimated along with the dynamical states, thanks to the high observability. In addition, the time-derived ranging method provides sufficient information to the filter in order to meet the navigation requirements. In these simulations, only the initial phase of the operative orbit has been considered. Basically, after one orbital period, it is quite expected to improve the estimation due to relative position change between \ac{S/C}. This would bring additional information to the filter and decrease the uncertainty. This study considered only the first 14 days of the mission. Even though this is sufficient to see the expected performances, further work would be needed to investigate what would happen for the full operative mission lifetime (1-year). This study also didn't consider the dynamic model errors and the effects of clock drift on the performances and they are considered as topics for future research.

\bibliographystyle{Frontiers-Harvard} 
\bibliography{articleFrontiers}

\end{document}